# Heavy-light charm mesons spectroscopy and decay widths

[1]Alka Upadhyay, [1]Meenakshi Batra, [1]Pallavi Gupta

[1]School of Physics and Material Science,

Thapar University Patiala (Punjab)-147004

## Abstract

We present the mass formula for heavy-light charm meson for one loop, using heavy quark effective theory. Formulating an effective Lagrangian, the masses of the ground state heavy mesons have been studied in the heavy quark limit including leading corrections from finite heavy quark masses and nonzero light quark masses using a constrained fit for the eight equation having eleven parameters including three coupling constants g, h and g'. Masses determined from this approach is fitted to the experimentally known decay widths to estimate the strong coupling constants, showing a better match with available theoretical and experimental data.

## 1. Introduction

A heavy hadronic system contains heavy quark with spin quantum number ($S_Q$) and light degrees of freedom where light degrees of freedom include light quark and gluons interacting through quark-antiquark pairs. The light degrees of freedom should have the quantum number of light quark that is $S_l$ in order to have total conserved quantum number J where $J=S_Q+S_l$. Defining J as $J^2=j(j+1)$ and $S_Q^2=s_Q(s_Q+1)$ and $S_l^2=s_l(s_l+1)$, the total spin $j_\pm = s_l \pm 1/2$ can be obtained by combining the spin of heavy quark spin ½ with spin of light degrees of freedom. The ground state l=0 heavy mesonic system form a degenerate doublet with j=0⊕1 and negative parity denoted as D and $D^*$ for charm meson. The first exited states are $0^+$ and $1^+$ with $D_0$ and $D_1$. The lowest lying excited states $J^P = 0^+$ and $1^+$ heavy mesons are the members of the $j^p = \frac{1^+}{2}$ doublet. There is also an excited doublet of heavy mesons with $j^p = \frac{3^+}{2}$, whose members are $J^P = 1^+$ and $2^+$. The members of various doublets are represented by : (P,P$^*$) for l=0, (P$_0^*$,P$_1^{'}$) for $J^P$=(0$^+$,1$^+$) while those with $J^P$=(1$^+$,2$^+$) for l=1 corresponds to (P$_1$,P$_2^*$) .

Study of heavy meson spectrum is one of the recent motivations for researchers. There are several resonances like D$_J$(3000)[1], D$^*_{SJ}$ (2860), D$_{SJ}$ (3040) [2] which still needs confirmation regarding their $J^P$ states. Heavy quark effective theory (HQET) is one of the effective tool to study details of recent resonances observed at different experiments. The theory can be applied to estimate the various signatures like masses, decay widths and their quantum numbers of meson states. Various models like quark models [3-4], potential models and lattice studies were used to calculate meson masses earlier times but the calculated charm masses were found to be of higher values as compared with experiments[5-10]. The approximate symmetries of QCD for heavy quarks can be incorporated to get information about the heavy-light system of mesons like charm and bottom mesons. In the infinite heavy quark mass limit, a heavy light system $Q\bar{q}$ can be classified into doublets depending upon their quantum numbers.

Heavy meson spectroscopy can be analysed by electron collider experiments but observations of such a long range of resonances lead to different puzzles. One of the most important evidence of $c\bar{s}$ broad states was provided by CLEO Collaboration [6], which observed a state of mass 2460 MeV and width 290 MeV. The BaBar Collaboration observed another narrow peak in the D$^+_s \pi^0$ invariant mass distribution, corresponding to a state of mass 2.317GeV [7]. Both of these states were confirmed by Focus and Belle collaborations [8]. Recent experimental evidence lies with Ds$_J$(2860) observed by BaBar Collaboration [9] with mass as 2856.6 $\pm$ 1.5 $\pm$ 50MeV and Belle Collaboration measured another peak in B$^+$ decays with mass as M(Ds$_J$ (2715)) = 2715 $\pm$ 11$^{+11}_{-14}$MeV [7]. Moreover, in the DK mass distribution, BaBar Collaboration noticed a broad structure with mass, M = 2688 $\pm$ 4 $\pm$ 3 MeV [10] and width $\Gamma$ =112•$\pm$7$\pm$36 MeV, likely the same resonance Ds$_J$(2700) found by Belle. Most recent state in the strange sector was announced in 2009 as Ds$_J$(3040) with mass M = 3044 $\pm$ 8(stat)+$^{30}_{-5}$(sys) MeV/c$^2$ [11]. This state was observed in D$^*$K channel mode not DK bound state. Many of the excited charm mesonic states in the non-strange sector have been recently analyzed [12] and needs more data with more accuracy. Even the experimental searches include 0$^-$ and 1$^-$ states in the non-strange sector but the excited states like 2$^+$ and 3$^+$ non-strange mesons have not yet been confirmed accurately.

In this paper, section 2 presents a heavy quark effective theory for writing an effective Lagrangian with leading order corrections from finite heavy quark masses and nonzero light quark masses. In section 3, we formulate the mass formulae for ground states and low lying excited states charm meson in terms of eleven unknown parameters and their decay widths in terms of strong coupling constants. Section 4 , we present the results for charm masses, decays and coupling constants and the constrained fits for the respective parameters and further conclusion.

## 2. Theoretical framework: Heavy Quark effective theory

The properties of hadrons with a heavy quark coupled with light degrees of freedom can also be explained on the basis of symmetry occurring in heavy quark limit which takes a particular simpler form in the limit $m_Q \to \infty$. Two theories related to two symmetries, chiral symmetry for light quarks u, d, s and heavy quark spin and flavor symmetry for heavy quarks c and b can be exploited to explain a system with one heavy quark and other light one. Heavy light mesons can be studied by implementing both the chiral symmetry and heavy quark spin and flavor symmetry in the form of an effective Lagrangian [13,24,25]. Effective Lagrangian here describes the interplay between the chiral symmetry and heavy quark symmetry in the form of low energy gradients with heavy and light fields as operators. For Heavy-light system, heavy quark doublets are represented by effective fields and the octet of light pseudo Goldstone bosons are grouped in a single field. The Chiral Lagrangian for heavy mesons incorporating the heavy quark and vector symmetry can be written by including a kinematic term and all the possible interactions with the Goldstone bosons which involve the symmetry breaking and conserving terms. The expressions for effective fields describing the various doublets in heavy quark limit are described as below [14].

$$H_a = \frac{1+\slashed{v}}{2}(P_a^{*\mu}\gamma_\mu - P_a\gamma_5) \qquad (1)$$

$$S_a = \frac{1+\slashed{v}}{2}(P_{1a}^{'\mu}\gamma_\mu\gamma_5 - P_{0a}^*) \qquad (2)$$

Here $a = u, d, s$ is the SU(3) index and various operators annihilate mesons of four velocity v.

Equation (1) and (2) represents ground state negative parity doublet($0^-, 1^-$) and low lying excited state positive parity doublet($0^+, 1^+$) respectively. The interactions among heavy and light mesons are obtained by expanding the field $\xi = e^{i\Pi/f}$ and $\Sigma = \xi^2$ where f is the pion decay constant. The pion octet is here introduced by the vector and axial combinations $V^\mu = \frac{1}{2}(\xi \partial^\mu \xi^\dagger + \xi^\dagger \partial^\mu \xi)$ and $A^\mu = \frac{1}{2}(\xi \partial^\mu \xi^\dagger - \xi^\dagger \partial^\mu \xi)$ as

$$\begin{bmatrix} \frac{1}{\sqrt{2}}\pi^0 + \frac{1}{\sqrt{6}}\eta & \pi^+ & K^+ \\ \pi^- & -\frac{1}{\sqrt{2}}\pi^0 + \frac{1}{\sqrt{6}}\eta & K^0 \\ K^- & \overline{K}^0 & -\frac{2}{\sqrt{6}} \end{bmatrix} \quad (3)$$

These fields are involved to form an effective Lagrangian.

Now the Lagrangian that describes the dynamics of mesons with the heavy-light combination is given by [15]

$$L = L_{kinetic} + L_{axial} + L_{ct}. \quad (4)$$

The kinetic part of the Lagrangian

$$L_{kinetic} = -Tr[\overline{H^a}(iv.D_{ba} - \delta_H \delta_{ab})H_b] + Tr[\overline{S^a}(iv.D_{ba} - \delta_S \delta_{ab})S_b], \quad (5)$$

Where δH and δS are the residual masses of H and S fields where the residual masses are defined to be the difference between the real mass and an arbitrarily chosen reference mass of $O(m_q)$. The axial coupling of the field to the pseudo-goldstone bosons is defined by the Lagrangian :

$$L_{axial} = gTr[\overline{H_a}H_b A_{ba}\gamma_5] + g'Tr[\overline{S_a}S_b A_{ba}\gamma_5] + hTr[\overline{H_a}S_b A_{ba}\gamma_5 + h.c.] \quad (6)$$

Where g, g' are coupling constants in the ground state and in exited state doublets respectively, and h is he coupling between mesons belonging to different doublets. The mass counter-term Lagrangian:

$$L_{ct} = Tr[(a_H \overline{H_a}H_b - a_S \overline{S_a}S_b)(\xi m_q \xi + \xi^\dagger m_q \xi^\dagger)_{ab}] + Tr[(\sigma_H \overline{H_a}H_a - \sigma_S \overline{S_a}S_a)(\xi m_q \xi + \xi^\dagger m_q \xi^\dagger)_{bb}], \quad (7)$$

Where $m_q$= diag($m_u$, $m_d$, $m_s$), $\xi^2 = \exp(2i\phi/f)$ with $\phi$ being usual matrix of pseudo-Goldstone bosons and f≈130MeV. In terms of heavy quark symmetry conserving and symmetry violating terms the above Lagrangian can be written as

$$L_v^{ct} = -\frac{\Delta_H}{8}Tr[\overline{H}_a\sigma^{\mu\nu}H_a\sigma_{\mu\nu}] + \frac{\Delta_s}{8}Tr[\overline{S}_a\sigma^{\mu\nu}s_a\sigma_{\mu\nu}] + a_H Tr[\overline{H}_a H_b]m_{ba}^\xi - a_s Tr[\overline{S}_a S_b]m_{ba}^\xi$$

$$+\sigma_H Tr[\overline{H}_a H_a]m_{bb}^\xi - \sigma_s Tr[\overline{S}_a S_a]m_{bb}^\xi - \frac{\Delta_H^{(a)}}{8}Tr[\overline{H}_a\sigma^{\mu\nu}H_b\sigma_{\mu\nu}]m_{ba}^\xi + \frac{\Delta_S^{(a)}}{8}Tr[\overline{S}_a\sigma^{\mu\nu}s_b\sigma_{\mu\nu}]m_{ba}^\xi \ldots\ldots(8)$$

$$-\frac{\Delta_H^{(a)}}{8}Tr[\overline{H}_a\sigma^{\mu\nu}H_a\sigma_{\mu\nu}]m_{bb}^\xi + \frac{\Delta_S^{(a)}}{8}Tr[\overline{S}_a\sigma^{\mu\nu}s_a\sigma_{\mu\nu}]m_{bb}^\xi,$$

where. Here $\Delta_H$, $\Delta_S$ terms in above equation are symmetry (spin) violating operators giving rise to hyperfine splitting and $a_H$, $a_S$, $\sigma_H$, $\sigma_S$ terms preserve spin-symmetry while other operators violate heavy quark spin symmetry.

### 3. Mass formula and decay width for charm mesons

In the framework of heavy hadron chiral perturbation theory chiral corrections and corrections due to chiral and heavy quark symmetry are encountered and at one loop level, the residual masses can be given by a generalized formula

$$m_{R_a}^0 = \delta_R + \frac{n_J}{4}(\Delta_R + \Delta_R^\sigma \overline{m} + \Delta_R^{(a)} m_a) + \sigma_R \overline{m} + a_R m_a + \frac{g_R^2}{f^2}c^{R_a}K_1(\eta,m) + \frac{h^2}{f^2}c^{R_a}K_2(\eta,m)\ldots\ldots(9)$$

Where R is an index that labels the ground state (H) and excited state (S), each of the ground state and excited states having members corresponding to J=0,1 where $n_J = n_0 = -3$ and $n_1 = 1$ These coefficients come from $S_Q . S_l$ operator and gives -3/4 for pseudoscalar mesons and ¼ for vector mesons. Here $m_a$=u, d, s and $\overline{m} = m_u + m_d + m_s$. . Here g, g' and h are the coupling constants corresponding to H-H, S-S and H-S interactions respectively. Thus in total we obtain 12 equations representing the residual masses at one loop level for low lying doublet combining with three light quarks respectively but assuming isospin symmetry the equations reduce to eight and $c^{R_a}$ are the coefficients that signifies the emission of pion, kaon and eta goldstone boson from a specific charm states. The index (a) labels the light flavor and runs over u, d, s and $K_1$ and $K_2$ are the chiral loop functions defined as,

$$K_1(\eta, M) = \frac{1}{16\pi^2}\left[(-2\eta^3 + 3M^2\eta)\ln\left(\frac{M^2}{\mu^2}\right) + 2\eta(\eta^2 - M^2)F\left(\frac{\eta}{M}\right) + 4\eta^3 - 5\eta M^2\right] \quad (10)$$

$$K_2(\eta, M) = \frac{1}{16\pi^2}\left[(-2\eta^3 + M^2\eta)\ln\left(\frac{M^2}{\mu^2}\right) + 2\eta^3 F\left(\frac{\eta}{M}\right) + 4\eta^3 - \eta M^2\right] \quad (11)$$

And

$$F(x) = 2\frac{\sqrt{1-x^2}}{x}\left[\frac{\pi}{2} - \text{Tan}^-\left(\frac{x}{\sqrt{1-x^2}}\right)\right] \qquad |x| < 1$$

$$F(x) = -2\frac{\sqrt{x^2-1}}{x}\ln(x + \sqrt{x^2 - 1})|x| > 1 \qquad (12)$$

The function $K_1(\eta, M)$ appears whenever the virtual heavy meson inside the loop is in the same doublet as the external heavy meson, while $K_2(\eta, M)$ appears when the virtual heavy meson is from the opposite parity doublet.

Here $\eta$ is the mass difference of heavy-light charm mesons in the initial and final state and M is the mass of the pseudo scalar boson which is emitted during the process. $\mu$ is the energy scale of 1 GeV.

Two body strong decays of heavy-light mesons to pseudo-scalar mesons can be derived from the heavy meson chiral lagrangian $L_0, L_{HH}, L_{SH}, L_{TH}$ etc, which can be written as:

$$L_0 = iTr\{\overline{H}_a v.D_{ab} H_b\} + iTr\{\overline{S}_a v.D_{ab} S_b\} + iTr\{\overline{T}_a^\mu v.D_{ab} T_{\mu b}\}$$
$$L_{HH} = gTr\{\overline{H}_a H_b \gamma_\mu \gamma_5 A_{ba}^\mu\},$$
$$L_{SH} = g'Tr\{\overline{H}_a S_b \gamma_\mu \gamma_5 A_{ba}^\mu\} + h.c.,$$
$$L_{SH} = hTr\{\overline{S}_a S_b \gamma_\mu \gamma_5 A_{ba}^\mu\} + h.c. \quad (13)$$

Here $D\mu$ is the covariant derivative and $A\mu$, $V\mu$ have their usual meanings $\Lambda$ is the chiral symmetry breaking to be taken as ~1 GeV, g,g',h, $g_{TH}$ are the hadronic coupling constants between the H-H, S-S, S-H, T-H state interaction respectively. The coupling constants depend on the radial quantum number of the heavy mesons and are represented by g~. From these chiral Lagrangian terms, we obtain decay width for strong interaction to final states $D^{(*)}\pi$, $D^{(*)}K$, $D^{(*)}\eta$ is then obtained as

$$\Gamma = \frac{1}{2J+1}\sum \frac{p_f}{8\pi M_i^2}|A|^2; \quad \text{where} \quad p_f = \frac{\sqrt{(M_i^2 - (M_f + m_p)^2)(M_i^2 - (M_f - m_p)^2)}}{2M_i} \quad (14)$$

Where $A$ is the scattering amplitude, i and f denotes the initial and final particles, J is the total angular momentum of the initial heavy meson, P denotes the light pseudo scalar meson, summation $\sum$ is of all the polarisation vectors, $p_f$ is the momentum exchanged between the initial and final state. The expressions for the decay widths for various doublets are as follow where M refers to the emitted pseudo-scalar meson ($\pi$, $\eta$, K) fields.

$(0^-, 1^-) \, to \, (0^-, 1^-) + M$

$$\Gamma(1^- \to 0^-) = c_P \frac{g_H^2 M_f p_f^3}{6\pi f_\pi^2 M_i} \quad (15)$$

$$\Gamma(1^- \to 0^-) = c_P \frac{g_{HH}^2 M_f p_f^3}{6\pi f_\pi^2 M_i} \quad (16)$$

$$\Gamma(1^- \to 1^-) = c_P \frac{g_{HH}^2 M_f p_f^3}{3\pi f_\pi^2 M_i} \quad (17)$$

$$\Gamma\left(0^- \to 1^-\right) = c_P \frac{g_{HH}^2 M_f p_f^3}{2\pi f_\pi^2 M_i} \tag{18}$$

$(0^+, 1^+)$ to $(0^-, 1^-) + M$

$$\Gamma\left(1^+ \to 1^-\right) = c_P \frac{g_{SH}^2 M_f \left(p_f^2 + m_p^2\right) p_f}{2\pi f_\pi^2 M_i} \tag{19}$$

$$\Gamma\left(0^+ \to 0^-\right) = c_P \frac{g_{SH}^2 M_f \left(p_f^2 + m_p^2\right) p_f}{2\pi f_\pi^2 M_i} \tag{20}$$

The coefficients $C_p$ are different for different pseudo-scalar mesons,

$$C_{\pi^\pm} = C_{K^\pm} = C_{K^0} = C_{\bar{K}^0} = 1, C_{\pi^0} = \frac{1}{2}, C_\eta = \frac{1}{6}$$

## 4. Result and Conclusion

Within heavy hadron chiral perturbation theory framework, the charm mesons masses, decay width and couplings are analysed using a mass formulae up to one loop chiral corrections [17]. With higher loop corrections at the order $O(Q^3)$ the mass formula in equation (9) consist of eight equations having eleven parameters. The calculations here depend on eleven parameters $\delta_s + \sigma_S \bar{m}, \delta_H - \sigma_H \bar{m}, a_H, a_S, \Delta_H^{(a)}, \Delta_S^{(a)}, g, g\,', h, \Delta_H + \Delta_H^{(\sigma)} \bar{m}, \Delta_S + \Delta_S^{(\sigma)} \bar{m}$.

We will use the experimentally measures residual masses written below, for fitting our eleven parameter. Two of the parameters, the axial coupling for the ground state doublet of charmed mesons, and h which dominates the strong coupling between even parity with the ground state charmed meson, given a range from 0-1.These constrains are justified from the literature [13]. The experimentally measured residual masses in reference to the non-strange spin averaged mass $(m_{H_1} + 3/4\, m_{H_1^*})$ are:

$$m_{H_1} = -106.1\, MeV \quad m_{H_3} = -4.75\, MeV \quad m_{H_1}^* = 35.4\, MeV \quad m_{H_3}^* = 139.1\, MeV \tag{21}$$

$$m_{S_1} = 335.0\, MeV \quad m_{S_3} = 344.4\, MeV \quad m_{S_1}^* = 465.0\, MeV \quad m_{S_3}^* = 486.3\, MeV$$

The tree level mass formula where g, g' and h=0, the residual masses can be reproduced with the many sets of parameters. There are large uncertainties in the parameters involve in the Lagrangian of the low lying excited states i.e even parity mesons. Below is one such set of eight parameters reproducing above residual masses at tree level.

$$\delta_s + \sigma_S \overline{m} - \delta_H - \sigma_H \overline{m} = 432 \pm 26 MeV,$$
$$\Delta_H + \Delta_H^{(\sigma)} \overline{m} = 146 \pm 1 MeV, \Delta_S + \Delta_S^{(\sigma)} \overline{m} = 129 \pm 50 MeV, \quad (22)$$
$$a_H = 1.14 \pm 0.06, a_S = 0.21 \pm 0.29, \Delta_H^{(a)} = -0.03 \pm 0.01, and\ \Delta_S^{(a)} = 0.14 \pm 0.55$$

In the mass formulae for charm meson, we can have otherwise 12 equations for all the masses of even and odd parity charm mesons depending on the light quark being u,d or s. But as soon as we impose the chiral symmetry, these 12 equation reduces to 8 mass equations in terms of 11 degrees of freedom. The importance of these parameters lies in their ability to explain the spin violation and flavor breaking at the higher loop calculations. Here some of the parameters $\sigma_H, \sigma_S, \Delta_H^{(\sigma)}, \Delta_S^{(\sigma)}$ cannot be separately determined, because they always appear in linear combination with other parameters like $\delta_H, \delta_s, \Delta_H, \Delta_S$ respectively. To make the calculation simpler, we will absorb the measured values of $\sigma_H, \sigma_S, \Delta_H^{(\sigma)}, \Delta_S^{(\sigma)}$ in $\delta_H, \delta_s, \Delta_H, \Delta_S$ respectively.

Different predictions from relativistic and non-relativistic quark model, and decays of D* in one loop calculation, restrict the value of coupling constants g and h to lie between 0 and 1 but $g'\epsilon[-1,1]$[16]. Using the literature value, we vary these values over the range 0-1. We used f = 120 MeV extracted in Ref. [13,17] using the one loop formulae for Pion and Kaon decay constants. We set $m_u = m_d = 4$ MeV and $m_s = 90$ MeV. Using Mathematica7.0 [18] as a programming language to fit the values, a number of sets can be obtained. The sets can be reduced by putting justified constrain on the unknown parameters. We cannot make strong conclusion on the sets, as there are some of the parameters, which cannot be significantly determined, and thus the uncertainty on parameters are very large. Using some of the constrains from literature on spin breaking and flavor breaking terms, values of couplings, one such set, that satisfies our residual experimental masses i.e equation (21) are given below. One of the best fitted parameters set to these residual masses is

$$g = 0.1, g' = 0.1, h = 0.07, \delta_H = 4, \delta_S = 431, \Delta_H = 144, \Delta_S = 126$$
$$a_H = 1.1, a_S = 0.21, \Delta^{(a)}{}_H = -0.04, \Delta^{(a)}{}_S = 0.14 \quad (23)$$

Based on the above parameters, central masses of charmed meson for one loop corrections have been calculated and presented in table 1 column 5. Here real masses and decay widths have been compared with the experimental data taken from Particle Data Group [5]. $D_0^0$ Mass has also been measured at Belle [19] and is equal to 2308±36 similarly other masses are also available. We present our results, that are matching well with PDG data [5].

| S.No | State | Mesonic State | $J^P$ | Calculated Residual Mass (MeV) | Calculated Real Mass (MeV) | Experimental Real Mass (MeV)[5] | Decay Width($\Gamma$) (MeV) |
|---|---|---|---|---|---|---|---|
| 1. | $m_{H_1}$ | $D^{0,+}$ | $0^-$ | $-105.97$ | 1867.04 | 1869.61 | |
| 2. | $m_{H_3}$ | $D_S^\pm$ | $0^-$ | $-5.27028$ | 1967.74 | 1968.30 | |
| 3. | $m_{S_1}$ | $D_0^0$ | $0^+$ | 329.722 | 2314.08 | 2318 | 267 |
| 4. | $m_{S_3}$ | $D_{s0}^+$ | $0^+$ | 341.077 | 2314.55 | 2317.7 | < 3.8 |
| 5. | $m_{H_1^*}$ | $D^{*0,+}$ | $1^-$ | 44.4608 | 2017.24 | 2010.26 | 0.0834 |
| 6. | $m_{H_3^*}$ | $D_S^{*+}$ | $1^-$ | 136.485 | 2109.49 | 2112.1 | <1.9 |
| 7. | $m_{S_1^*}$ | $D_1^0$ | $1^+$ | 460.509 | 2433.52 | 2421.4 | 27.4 |
| 8. | $m_{S_3^*}$ | $D_{s1}^+$ | $1^+$ | 483.281 | 2456.29 | 2459.5 | <3.5 |

Table 1: Ground state and excited state even and odd parity charm mesons with their residual and real masses and decay widths.

Here the importance of results can be studied well from the calculated data. The data is obtained for the higher loop correction into the mass formulae (9). Spin symmetry violating parameters $\Delta_H$ and $\Delta_S$ for the ground and the excited states resulting from ($1/m_Q$) corrections in effective Lagrangian breaks the heavy quark spin symmetry. This results in the breaking of mass degeneracy between the members of the same doublets. These ($1/m_Q$) corrections will introduce the hyperfine splittings of the order 140MeV. The column 2 of the Table 2 shows the hyperfine splittings for our calculated masses and in the next column these splittings have been compared with the experimental values. This comparison shows that our calculated values are within the 2% deviation from the experimental values.

The SU(3) symmetry breaking arise when $m_u = m_d \neq m_s$ in the chiral Lagrangian at one loop mass formulae. This flavour breaking between the strange and non-strange charm mesons, whose other quantum numbers are identical, is expected to be of the order of ~100MeV corresponding to the mass of strange quark. The SU(3) flavor violating terms are entering in the form of $a_H$ and $\Delta_H^{(a)}$. Due to the eleven parameter fit, the constrains on the SU(3) flavor violating terms are showing unusual patterns but are with in 2-13% deviation from the experimental data.

| | Spin Splitting | | | Mass splitting | |
|---|---|---|---|---|---|
| State $J^P$ | Value from Experiment (MeV) | Calculated Value (MeV) | State | Value from Experiment (MeV) | Calculated Value (MeV) |
| $D_s^* - D_s(1^+ - 0^-)$ | 143.8 | 141.4 | $D_s^* - D_u^*(1^- - 1^-)$ | 105.4 | 94.04 |
| $D_s^* - D_s(1^+ - 0^+)$ | 141.9 | 138.7 | $D_s - D_u(0^+ - 0^+)$ | 98.8 | 96.6 |
| $D_u^* - D_u(1^- - 0^-)$ | 142.1 | 143.9 | $D_s^* - D_s^*(1^+ - 1^+)$ | 21.3 | 20.3 |
| $D_u^* - D_u(1^+ - 0^+)$ | 130 | 126.5 | $D_s - D_u(0^+ - 0^+)$ | 9.4 | 8.1 |

Table 2: Experimental and calculated value for spin splitting and mass splitting of meson doublets.

The decays channel of radially excited strange charm meson state [20] are used to justify our calculated masses using the effective theory. Here we calculate the strong coupling constant and verified it with other measures value in Ref[21].

$D^*_0$ (2700) → $(1^- → 0^-)$ DK; $(1^- → 1^-)$ D*K; $(1^- → 0^-)$ D$_S$ η; $(1^- → 1^-)$ D$_S$*η;

D* K$_0$*(this mode is excluded in our calculation as it gives negative momentum of light meson). Using the total decay width (125MeV experimentally) we get g~ = 0.30 which can be compared with g~=0.28 [21]. Decay width of strange $0^+$ and $1^+$ is used from [3] and coupling constant is determined and matched with h=-0.6[3].

| Decay mode | Width used (MeV)[22] | Coupling constant h calculated |
|---|---|---|
| D$_s$* → D$_s$Π $(0^+ → 0^-)$ | 260 | 0.7 |
| D$_s$ → D$_s$Π $(0^+ → 0^-)$ | 160 | 0.5 |

To check the validity of our approach we took the two different values of strong coupling i.e h=0.7 and h=0.5 and find the decay width for non-strange P (L=1) wave meson states. The results of the decay width are compared and shown to be matching well with the experimental values.

| Decay mode | Width(MeV): Calculated; h=0.7 | Width(MeV): Calculated; h=0.5 | Width (MeV) |
|---|---|---|---|
| D* → DΠ | 236.76 | 120.79 | 267 [5] |
| D → D*Π | 101.73 | 51.905 | 84[23] |

The above mentioned calculations can be summarised as, in this work, experimental data of ground state charmed meson masses is used to constrain eleven parameters in the effective Lagrangian to compute the masses, including higher loop corrections. The coupling constants and the mass parameters in effective Lagrangian reproduce QCD in specific limits and represent important input parameters for the description of the hadrons properties like heavy-light charm meson masses and mass splitting. The results are obtained and matched with the available experimental and theoretical data. Also to check the accuracy of the observed residual masses, these calculated masses have been used in decay width formulas and coupling constant (h) and decay width of excited state is calculated and it matches with various available theoretical and experimental information